%%    Title: 	Distributional curvature of time-dependent cosmic
%%    		strings (Preprint) 
%%
%%    Author: 	Jonathan Wilson
%%    Date:	13 May 1996
%%

% This paper has been submitted to Classical and Quantum Gravity,
% therefore the original version of this file used IOP macros. This
% file is a totally free-standing Plain TeX version which requires no
% additional macro files.

\font\bigbf=cmbx10 scaled \magstep2

% Use A4 paper

\hsize=160truemm
\vsize=235truemm
\hoffset=0truemm
\voffset=0truemm
\parskip=\smallskipamount

% Macros for section numbering etc. (For compatibility with IOP
% macros) 
%
\newcount\secno
\newcount\subno
\def\section#1{%
    \subno=0\global\advance\secno by 1 \bigbreak\leftline{\bf
    \number\secno. #1}\medbreak\noindent\ignorespaces}
\def\subsection#1{\global\advance\subno by 1
     \medbreak\leftline{\sl \number\secno.\number\subno. #1}\par
     \noindent\ignorespaces}
\def\definition#1{\bigbreak\noindent{\sl Definition #1}.%
  \quad\ignorespaces}
\def\Proof{\bigbreak\noindent{\sl Proof}.\quad\ignorespaces}
\def\endproclaim{\par\bigbreak}
\let\Bbb=\bf
\let\bdi=\bf
\def\d{{\rm d}}
\def\e{{\rm e}}
\def\i{{\rm i}}
\def\etal{{\sl et al}\ }
\def\scri{\three{\cal I}}
\def\fl{}

% Additional definitions

\catcode`\@=11
%
% Set abstraction.
%
\def\setabst#1#2{%
 \mathchoice{\set@bst\displaystyle{#1}{#2}}
            {\set@bst\textstyle{#1}{#2}}
            {\set@bst\scriptstyle{#1}{#2}}
            {\set@bst\scriptscriptstyle{#1}{#2}} }
\def\set@bst#1#2#3{{%
 \setbox0=\hbox{$#1#2#3$}
 \setbox1=\vbox to\ht0{}
 \mathopen{\hbox{$\left\{\copy1\right.\n@space$}} \, #2
 \mathrel{\hbox{$\left|\copy1\right.\n@space$}} #3 \,
 \mathclose{\hbox{$\left\}\copy1\right.\n@space$}}}}%
%
% Use \displaystyle entries in \cases
%
\def\cases#1{\left\{\,\vcenter{\normalbaselines\openup1\jot\m@th
    \ialign{$\displaystyle{##}\hfil$&\quad##\hfil\crcr#1\crcr}}\right.}
%
% multiple \eqalign
%
\def\multeqalign#1{\null\,\vcenter{\openup1\jot \m@th
  \ialign{\strut\hfil$\displaystyle{##}$&&$\displaystyle{{}##}$\hfil
  &\quad\hfil$\displaystyle{##}$\crcr#1\crcr}}\,}

\def\crssk{\cr\noalign{\smallskip}}
\def\Proposition{\bigbreak\noindent{\it Proposition}.%
  \quad\ignorespaces}
%
% other symbols etc.
%
\def\tfrac#1#2{{\textstyle{#1\over#2}}}
\def\norm#1{\left|#1\right|}	
\def\dirac#1{\delta^{(#1)}}	
\let\implies=\Longrightarrow    
\let\eps=\varepsilon            
\def\Real{{\Bbb R}}
\def\derv#1{\partial/\partial#1}
\def\dbyd#1#2{{\d#1\over\d#2}}
\def\pbyp#1#2{{\partial#1\over\partial#2}}
\def\supp{\mathop{\rm supp}\nolimits}
\def\dr{\d r}
\def\ds{\d s}
\def\dt{\d t}
\def\du{\d u}
\def\dx{\d x}
\def\dy{\d y}
\def\dz{\d z}
\def\duxyz{\,\du\,\dx\,\dy\,\dz}
\def\cA{{\cal A}}
\def\cD{{\cal D}}
\def\cE{{\cal E}}
\def\cG{{\cal G}}
\def\bE{{\bdi E}}
\def\bx{{\bdi x}}
\def\tg{{\tilde g}}

\def\tT{\tilde T}
\def\RCD#1#2#3{\left.{\tilde R}^{#1}{}_{#2}%
  \right._{#3}\sqrt{-\tg_{#3}}}
\def\RDD#1#2{{R^{#1}{}_{#2}\sqrt{-g}}}
\def\tmet#1{\widetilde{g_{#1}}_\eps}
\def\clap#1{\hbox to0pt{\hss#1\hss}}
\def\three#1{\mskip2mu{}^{\scriptscriptstyle(3)}\mskip-2mu#1}
\def\threeh#1{\mskip2mu{}^{\scriptscriptstyle(3)}\mskip-2mu\hat{#1}}
\def\iii{{\scriptscriptstyle(3)}}
\def\gg{\three{g}}
\def\gT{\three{T}}
\def\hr{\hat r}
\def\hm{\hat m}
\def\hn{\hat n}
\def\hL{\threeh L}
\def\hV{\hat V}
\def\hS{\threeh S}
\def\hT{\threeh T}
\def\hg{\threeh g}
\def\hnabla{\threeh\nabla}
\def\ung#1{\underline{g_{#1}}}
\def\unL#1{\underline{L_{#1}}}
\def\unS#1{\underline{S_{#1}}}

\catcode`\@=12

\vglue\bigskipamount
\centerline{{\bigbf Distributional curvature of time dependent cosmic
strings}} 
\bigskip
\centerline{{\bf J.~P.~Wilson}}
\bigskip
\centerline{{\bf Department of Mathematics, University of Southampton,
Southampton, SO17 1BJ}} 
\bigskip\medskip
\noindent{\bf Abstract}\quad
Colombeau's theory of generalised functions is used to calculate the
contributions, at the rotation axis, to the distributional curvature
for a time-dependent radiating cosmic string, and hence the mass per
unit length of the string source.  This mass per unit length is
compared with the mass at null infinity, giving evidence for a global
energy conservation law.

\vskip2\bigskipamount

\section{Introduction}
An important problem in relativity is the calculation of
distributional curvatures of non-regular metrics. Such curvatures
represent physical fields whose support is confined to a proper
submanifold of space-time, such as the point mass in the Schwarzschild
solution and cosmic strings.

The main obstacle to rigorously calculating such curvatures is the
ambiguity that arises when one tries to multiply distributions,
because the Riemann tensor is a nonlinear function of the metric and
its derivatives. It was proved by Schwartz~(1954) that one cannot
define an associative multiplication of distributions which is
compatible with the usual operation of differentiation. There are
metrics for which the distributional Riemann tensor may be calculated
directly without the need to multiply distributions, but this class of
metrics is very small.  In fact a necessary condition for the metric
to lie in this class is that the distributional curvature must have
its support on a submanifold of codimension of at most one (Geroch and
Traschen,~1987); thus it is possible for metrics representing matter
shells to lie in this class, but not for strings and point particles.

In an earlier paper (Clarke \etal 1996) it was shown how one may
apply Colombeau's theory of Generalised Functions (Colombeau, 1983,
1990), in which nonlinear operations are well defined, to the
calculation of distributional curvatures where a direct calculations
would not work. In Colombeau's theory, the Schwartzian space of
distributions $\cD'(\Real^n)$ is embedded as a linear subspace of a
much larger differential algebra $\cG(\Real^n)$ by the smoothing
operation
$$ \tilde f(\Phi,\bx) = {1\over\eps^n} \int_{\Real^n} f(\bx')
   \Phi\left({\bx'-\bx \over \eps}\right) \,\dx', $$
where $\Phi\in\cD(\Real^n)$ has unit radius. Thus products of two
distributions are defined in this algebra, although one cannot
interpret this product as a distribution via the embedding. However
one may give a distributional interpretation to this object via the
notion of weak equivalence of two generalised functions $F$ and $G$
(written as $F\approx G$) which essentially is the generalisation of
the equality of two distributions to $\cG(\Real^n)$.

In the previous paper (Clarke \etal 1996), the focus was on
calculating distributional curvatures of 2-surfaces, in particular the
cone with a metric of
$$ \ds^2 = \dr^2 + A^2r^2\d\phi^2, \qquad |A|<1. $$
Here the Gaussian curvature density $\tfrac12 R \sqrt{g}$ (as a
generalised function) is weakly equivalent to the distribution
$2\pi(1-A) \dirac2(x,y)$, where $(x,y)=(r\cos\phi,r,\sin\phi)$.  In
this paper we shall consider applications to axisymmetric dynamical
space-times with conical singularities and in particular space-times
with a pure radiation energy-momentum tensor (Kramer \etal 1980). That
is with the energy-momentum tensor of the form
$$ T_{ab} = \Phi^2 k_a k_b, \qquad k_a k^a=0, $$
where vector $k^a$ is the null direction of propagation and $\Phi$
represents the amplitude. In the case of cylindrical symmetry one may
write the metric as 
$$ ds^2 = e^{-2\psi} \bigl( e^{2\gamma} (-dt^2+dr^2) + r^2 d\phi^2
   \bigr) + e^{2\psi} dz^2, $$
where $\psi$ and $\gamma$ are functions of $t$ and $r$. One may then
obtain a pure radiation solution with
$$ \eqalign{ & k_t = -k_r = 1 \cr & \Phi^2 = r^{-1} \gamma_{,r} -
   {\psi_{,t}}^2 - {\psi_{,r}}^2 \cr }$$
by solving the following field equations (Kramer \etal 1980;
Krishna Rao, 1964) 
$$ \eqalign{ & \psi_{,rr} + r^{-1} \psi_{,r} - \psi_{,tt} =0 \cr &
   \gamma_{,r} + \gamma_{,t} = r (\psi_{,r} + \psi_{,t})^2 \cr }$$

The Colombeau framework will be used to extract the distributional
contributions at the axis $r=0$ for the Riemann tensor densities
$\RDD{ab}{cd}$ and the corresponding
distributional  contributions for the energy-momentum tensor, which is
defined via the field equations
$$ R_{ab}-\tfrac12 R g_{ab} = T_{ab} $$
Such distributional contributions may then be used to give a
definition of mass per unit length of such a string.

Finally we shall consider the physical implications; the behaviour of
matter at null infinity, using a three dimensional asymptotic
formalism, developed be Ashtekar \etal (1997).  This will Demonstrate
the sense in which global conservation holds.

\section{Distributional curvature of pure radiation fields}
In this paper we shall focus on the distributional curvature of a
particular cylindrically symmetrical pure radiation solution
$(\psi=0,\gamma=\gamma(t-r))$, where $\gamma(u)$ is an arbitrary
$C^\infty$ function, whose metric may be expressed as
$$ \ds^2 = \e^{2\gamma(t-r)} (-\dt^2+\dr^2)+r^2\d\phi^2+\dz^2.
   \eqno(1)$$
We may write this metric in null Cartesian coordinates
$(u=t-r,x=r\cos\phi,y=r\sin\phi,z)$;
$$ \eqalign{
   \ds^2 &= -\e^{2\gamma(u)} \du^2 + \tfrac12 (\dx^2 + \dy^2) +\dz^2
   -2\e^{2\gamma(u)} \left( {x\over r} \du\,\dx + {y\over r} \du\,\dy
   \right) \cr
   &\quad - {1\over2} \left( {x^2-y^2\over r^2} \dx^2 +
   4{xy\over r^2} \dx \dy + {y^2 - x^2 \over r^2} \dy^2 \right).
}$$
Except at the singularity $r=0$, the Riemann curvature density will
have non-zero independent components
$$ \RDD{xy}{ux} = - \gamma'/r \sin\phi, \quad
   \RDD{xy}{uy} =   \gamma'/r \cos\phi. $$
which may of course be interpreted as distributions since such
functions will always be locally integrable over $\Real^2$.  This does
not however mean that this distributional curvature will be valid at
the axis; we could expect a delta function contribution to the
distributional curvature due to a non-zero angular deficit which may
be shown to be
$$ \lim_{r\to0} 2\pi\left\{ 1- \left( g_{\phi\phi;a}
   g_{\phi\phi}{}^{;a} \over 4  g_{\phi\phi}
   \right)^{1/2} \right\} = 2\pi(1-e^{-\gamma(t)}), $$
as in the case of the static cone (See Clarke \etal 1996). We shall be
using a regularisation based upon Colombeau's generalised functions to
recover the distributional contributions, on the axis, to the
curvature.

It may be remarked that this singularity is not quasi-regular but
intermediate; that is the Riemann tensor components do not converge in
the parallelly propagated orthonormal frame $\bE$ along the integral
curves of $\derv{r}$
$$ \eqalign{
   \bE_0 &= \tfrac12 (1+\e^{-2\gamma}) \pbyp{}{t} + \tfrac12
   (1-\e^{-2\gamma}) \pbyp{}{r} \cr
   \bE_1 &= \tfrac12 (1-\e^{-2\gamma}) \pbyp{}{t} + \tfrac12
   (1+\e^{-2\gamma}) \pbyp{}{r} \cr
   \bE_2 &= r^{-1} \pbyp{}{\phi} \cr
   \bE_3 &= \pbyp{}{z} \cr
}$$
in which
$$R_{0202}=-R_{0212}=R_{1212} = -\e^{-4\gamma} {\gamma' \over r}. $$
However there is an orthonormal frame $\bE'$ in which they do converge
$$ \eqalign{
   \bE'_0 &= {1+r^2 \over 2r} \e^{-\gamma} \pbyp{}{t} +
   {1-r^2 \over 2r} \e^{-\gamma} \pbyp{}{r} \cr
   \bE'_1 &= {1-r^2 \over 2r} \e^{-\gamma} \pbyp{}{t} + 
   {1+r^2 \over 2r} \e^{-\gamma} \pbyp{}{r} \cr
   \bE'_2 &= r^{-1} \pbyp{}{\phi} \cr
   \bE'_3 &= \pbyp{}{z} \cr
}$$
for which
$$R'_{0202}=-R'_{0212}=R'_{1212} = -r \e^{-2\gamma} \gamma'. $$

We shall expect our distributional curvature to have cylindrical
symmetry; this suggests that we should calculate $\tmet{ab}$ by
smoothing kernel $\Phi(u,x,y,z)\in\cA_1(\Real^4),$ with radius
$$ R_0 = \sup\setabst{(x^2+y^2)^{1/2}}{ \int_{\Real^2}
   \norm{\Phi(u,x,y,z)}\,\du\,\dz >0}, $$ 
The smoothed metric may then be written as
$$ \fl \tmet{ab} = {1\over\eps^4}\int_{\Real^4}
   g_{ab}(u'+u,x'+x,y'+y,z'+z)
   \Phi\bigl(u'/\eps,x'/\eps,y'/\eps,z'/\eps \bigr)
   \,\du'\,\dx'\,\dy'\,\dz'.$$
We shall expect our distributional curvature to have cylindrical
symmetry; this suggests that it should be sufficient to use a
cylindrically symmetrical smoothing kernel, which we shall denote as
$$ \Phi(u,r)=\Phi(u,r\cos\phi,r\sin\phi,z). $$
With such a kernel the smoothed metric becomes
$$ \fl \tmet{ab} = {1\over\eps^3}\int_{\Real^4}
   g_{ab}(u'+u,x'+x,y'+y)
   \Phi\bigl(u'/\eps,(x'^2+y'^2)^{1/2}/\eps\bigr)
   \,\du'\,\dx'\,\dy'.$$

The metric components will be in general sums of $C^\infty$ terms,
which may be identified with their smoothings in $\cE_M(\Real^4)$, and
singular components. The only functions that we will have to smooth
are
\item{(i)} $(u,x,y,z)\mapsto \e^{2\i\phi}$ \par
\item{(ii)} $(u,x,y,z)\mapsto \e^{2\gamma(u)+\i\phi} $.  \par
For (i)  the smoothing is given by (see Clarke \etal 1996)
$$ \eqalign{
   \widetilde{\e^{2\i\phi}} &= {1\over\eps^3} \int^\infty_{-\infty}
   \! \int^\infty_0 \!\!\! \int^{2\pi}_0 \left(
   {r\e^{\i\phi}+r'\e^{\i\phi'}  
   \over r\e^{\i\phi'}+r'\e^{\i\phi}} \right) \e^{\i(\phi+\phi')}
   \Phi(u'/\eps,r'/\eps) r'\,\du'\,\dr'\,\d\phi' \cr
   & = H_\eps(r) \e^{2i\phi} \cr
}$$
where
$$ H_\eps(r)=\cases{
   2\pi \int^\infty_{-\infty} \! \int^{r/\eps}_0 \left( 1-{\eps^2
   r'^2\over r^2} \right) \Phi(u',r') r' \du'\,\dr', & for $r<\eps$,
   \cr
   1-2\pi{\eps^2\over r^2} \int^\infty_{-\infty} \! \int^{R_0}_0 r'^3
   \Phi(u',r') \,\du' \,\dr', & for $r>\eps$. \cr
}$$
For (ii) the smoothing is given by
$$ \fl \widetilde{\e^{2\gamma(u)+\i\phi}}= {1\over\eps^3}
   \int^\infty_{-\infty} \! \int^\infty_0 \!\!\!\int^{2\pi}_0
   e^{2\gamma(u+u')} \left( {r\e^{\i\phi}+r'\e^{\i\phi'} \over
   r\e^{\i\phi'}+r'\e^{\i\phi}} \right)^{1/2} \!
   \e^{\i(\phi+\phi')/2} \Phi(u'/\eps,r'/\eps) r'
   \,\du'\,\dr'\,\d\phi'
$$ 
in which $\phi'$ may be integrated out by complex contour integration;
If 
$$ I=\int^{2\pi}_0 \left(r\e^{\i\phi}+r'\e^{\i\phi'} \over
   r\e^{\i\phi'}+r'\e^{\i\phi}\right)^{1/2} \! \e^{\i(\phi+\phi')/2}
   \d\phi',
$$ 
then setting $w=\e^{\i\phi}$ and $z=\e^{\i\phi'}$, we may integrate out 
$\phi'$ by integrating the complex function
$$ \Gamma(z)=-\i \left( r'z + rw \over rz + r'w \right)^{1/2}
   {w\over z}^{1/2} $$
around the circular contour $\kappa:\norm{z}=1$ to obtain
$$ I= \cases{ 4E(r'/r) e^{i\phi}, & for $r'<r$, \cr 4G(r/r')
   e^{i\phi}, & for $r'>r$ \cr} $$
where the hypergeometric functions $E(\lambda)$ and $G(\lambda)$ are
defined by
$$ \eqalign{
   E(s) &= \int^1_0 \left( 1-\lambda^2 s^2 \over 1-\lambda^2
   \right)^{1/2}\d\lambda \cr 
   G(s) &= s \int^1_0 \left( 1-\lambda^2 \over 1- \lambda^2
   s^2 \right)^{1/2} \d\lambda \cr
}$$
This enables us to  express $\widetilde{\e^{2\gamma(u)+\i\phi}}$ as
$L_\eps(u,r)\e^{2\gamma(u)+\i\phi}$, where
$$ \fl L_\eps(u,r)=\cases{
   4\int^\infty_{-\infty} \! \int^{r/\eps}_0 \e^{2(\gamma(u+\eps
   u')-\gamma(u))} E(\eps r'/r) \Phi(u',r') r' \du'\,\dr' \cr
   \quad + 4\int^\infty_{-\infty} \! \int^\infty_{r/\eps}
   \e^{2(\gamma(u+\eps u')-\gamma(u))}  
   G(r/\eps r') \Phi(u',r') r' \,\du'\,\dr', & for $r<\eps R_0$, \cr 
   4\int^\infty_{-\infty} \! \int^\infty_0 \e^{2(\gamma(u+\eps
   u')-\gamma(u))} E(\eps r'/r) \Phi(u',r') r' \,\du'\,\dr',&  for
   $r>\eps R_0$. \cr 
}$$
We may therefore express $\tmet{ab}$ as
$$ \fl \eqalign{
   \widetilde{\ds^2} &= -\e^{2\gamma(u)} \du^2 + \tfrac12 (\dx^2 +
   \dy^2) +\dz^2 -2\e^{2\gamma(u)} L_\eps(u,r) \left( {x\over r}
   \du\,\dx + {y\over r} \du\,\dy \right) \cr
   &\quad - {1\over2} H_\eps(r) \left( {x^2-y^2\over r^2} \dx^2 +
   4{xy\over r^2} \dx \dy + {y^2 - x^2 \over r^2} \dy^2 \right).
}$$
The smoothed metric will be used to calculate distributional
curvatures.  In almost all cases, the following estimates for
$H_\eps(r)$ and $L_\eps(u,r)$ will prove to be sufficient;
$$ \eqalign{
    H_\eps(r) &= \cases{ O(r^2/\eps^2), & for $r<\eps R_0$, \cr
    1+O(\eps^2/r^2), & for $r>\eps R_0$ \cr}  \cr
    L_\eps(u,r) &= \cases{ O(r^2/\eps^2), & for $r<\eps R_0$, \cr
    1+O(\eps^2/r^2), & for $r>\eps R_0$ \cr}  \cr
}$$
It must be emphasised that $L_\eps$ retains its dependence of $u$,
even though we shall omit it from the order notation; on the other
hand, $H_\eps$ is only dependent on $r$ and $\eps$.

We now calculate the generalised Riemann density functions
$\RCD{ab}{cd}{}$ from the smoothed metric $\tmet{ab}$;
$$ \multeqalign{
   \RCD{ux}{ux}\eps &= P_0 + P_1 \cos2\phi &&&
   \RCD{uy}{uy}\eps &= P_0 - P_1 \cos2\phi \cr
   &&& \clap{$\displaystyle{\RCD{ux}{uy}\eps} =
   \RCD{uy}{ux}\eps = P_1 \sin2\phi$}&& \cr
   \RCD{xy}{ux}\eps &= -P_2 \sin\phi &&&
   \RCD{xy}{uy}\eps &=  P_2 \cos\phi \cr
   \RCD{ux}{xy}\eps &=  L_\eps P_3 \sin\phi &&&
   \RCD{uy}{xy}\eps &= -L_\eps P_3 \cos\phi \cr
   &&&\clap{$\displaystyle{\RCD{xy}{xy}\eps} = P_3$}&& \cr
}$$
where
$$ \eqalign{
   P_0 &= \cases{ O(r/\eps^2), & for $r<\eps R_0$, \cr O(\eps^2/r^3),
   & for $r>\eps R_0$, \cr} \crssk
   P_1 &= \cases{ O(r/\eps^2), & for $r<\eps R_0$, \cr O(\eps^2/r^3),
   & for $r>\eps R_0$, \cr} \crssk 
   P_2 &= \cases{ -\gamma(u)/r + O(1/r), & for $r<\eps R_0$, \cr
   -\gamma(u)/r + O(\eps^2/r^3), & for $r>\eps R_0$, \cr} \crssk
  P_3 &= \cases{ O(1/\eps^2), & for $r<\eps R_0$, \cr O(\eps^2/r^4),
   & for $r>\eps R_0$, \cr} \cr
}$$

To obtain the distributional parts to the curvature on the axis not
contained in the expressions for $\RDD{ab}{cd}$, one must evaluate the
limits
$$ \lim_{\eps\to0} \int_{\Real^4} \left( \RCD{ab}{cd}\eps -
    \RDD{ab}{cd} \right) \Psi \duxyz $$
where $\Psi$ is an arbitrary function in $\cD(\Real^4)$. The following
notations will be used throughout the analysis;
$$ \eqalign{
   & K=\supp\Psi, \cr
   & K' = \setabst{(u,z)}{(u,x,y,z)\in K}, \cr
   & B_\eps = \setabst{(u,x,y,z)\in K}{(x^2+y^2)^{1/2} \leq \eps R_0},
   \cr
   & R_K = \sup \setabst{(x^2+y^2)^{1/2} }{(u,x,y,z)\in K}. \cr
}$$

We shall first show that $[\RCD{xy}{ux}\eps-\RDD{xy}{ux}]
\approx0$. That is
$$ \int_{\Real^4} \left( \RCD{xy}{ux}\eps - \RDD{xy}{ux} \right)
   \Psi(u,x,y,z) \duxyz \to 0. $$
We have
$$ \eqalign{
   &\norm{\int_{\Real^4} \left( \RCD{xy}{ux}\eps
   -\RDD{xy}{ux} \right) \Psi \duxyz} \cr 
   &\qquad \leq M_1 \int_{B_\eps} \norm{\RCD{xy}{ux}\eps
   -\RDD{xy}{ux}} \duxyz \cr
   &\qquad\quad + M_1 \int_{K-B_\eps} \norm{\RCD{xy}{ux}-\RDD{xy}{ux}}
   \duxyz \cr
   &\qquad \leq M_2 \eps + M_3 \left( {\eps^2 \over \eps R_0} -
  {\eps^2\over R_K} \right)
}$$
where we shall use $M_i$ to denote positive constants. Therefore
$$ \int_{\Real^4} \left(\RCD{xy}{ux}\eps - \RDD{xy}{ux} \right)
   \Psi(u,x,y,z) \duxyz =O(\eps). $$
Similarly it may be shown that 
$$ \eqalign{
   & [\RCD{xy}{u\alpha}{}] \approx
   \RDD{xy}{u\alpha} \cr 
   & [\RCD{u\alpha}{u\beta}{}] \approx 0. \cr
}\qquad \alpha,\ \beta=x,\ y$$

For the component $\RCD{ux}{xy}{}$ (and similarly for
$\RCD{uy}{xy}{}$), we may write (using the mean value theorem)
$$ \int_{\Real^4} \RCD{ux}{xy}\eps \Psi \duxyz = I_1 + I_2 $$
where
$$ \eqalign{
   I_1 &= \int_K \RCD{ux}{xy}\eps \Psi(u,0,0,z) \duxyz \cr
   I_2 &= \int_K \RCD{ux}{xy}\eps \dbyd{\Psi}{r} (u,\xi x,\xi y,z)r
  \duxyz \cr
   \xi & \in [0,1] \cr
}$$
now
$$ \eqalign{
   \norm{I_2} &\leq M_4 \int_{B_\eps} \norm{\RCD{ux}{xy}\eps } r
   \duxyz \cr
   &\quad + M_4 \int_{K-B_\eps} \norm{\RCD{ux}{xy}\eps} r \duxyz \cr
   & \leq M_5 {(\eps R_0)^3 \over \eps^2} + M_6 \eps^2\left(
  {1\over\eps R_0} - {1\over R_K}\right), \cr
}$$
and
$$ I_1 = \int_{K'} \left( \int_0^\infty \! L_\eps {P_3} r\,\dr
   \right) \left(\int_0^{2\pi}\! \sin\phi \,\d\phi\right)
   \Psi(u,0,0,z) \,\du\,dz
$$
which vanishes because our radial integral has no $\phi$-dependence
and is bounded. This will imply that
$$ \int_{\Real^4} \RCD{ux}{xy}\eps \Psi \duxyz = O(\eps) $$
and hence that
$$ \RCD{u\alpha}{xy}{} \approx 0 \qquad \alpha=x,\,y $$

Only one component now remains; $\RCD{xy}{xy}{}$, which we can expect
to have a delta function contribution to the distributional curvature
at the axis. We shall require a more accurate estimate for
$\RCD{xy}{xy}\eps$ derived from
$$ \RCD{xy}{xy}\eps = -\e^{\gamma} {\d\over\dr} \left[ 2\pi
    \int^{r/\eps}_0 \Phi(r') r'\,\dr + 1 \over (1+{H_\eps})^{1/2}
    (1-{H_\eps}+2\e^{2\gamma}{L_\eps}^2)^2 \right]
    $$
This will imply that
$$ \fl \int^{R_K}_0 \RCD{xy}{xy}\eps r\,\dr = \e^{\gamma} \left( 1-
   {2\pi \int^{{R_K}/\eps}_0 \Phi(r') r'\,\dr + 1 \over
   (1+{H_\eps({R_K})})^{1/2}
   (1-{H_\eps({R_K})}+2\e^{2\gamma}{L_\eps(u,{R_K})}^2)^2}\right).$$ 
But ${H_\eps}(0)={L_\eps}(u,0)=0$, $H_\eps({R_K})=1+O(\eps^2)$ and
$L_\eps(u,{R_K})=1+O(\eps^2)$ so
$$ \int^{R_K}_0 \RCD{xy}{xy}\eps r\,\dr = \e^\gamma (1-\e^{-\gamma})
   +O(\eps) $$

We now consider the full four dimensional integral;
$$ \int_{\Real^4} \RCD{xy}{xy}\eps \Psi \duxyz = I_3+I_4 $$
where for some $\xi\in[0,1]$,
$$ \eqalign{
   I_3 &= \int_K \RCD{xy}{xy}\eps \Psi(u,0,0,z) \duxyz, \cr
   I_4 &= \int_K \RCD{xy}{xy}\eps \dbyd{\Psi}{r}(t,\xi x,\xi
   y,z)r\duxyz, \cr
}$$
Now
$$ \eqalign{\norm{I_4} 
   & \leq M_7 \int_{B_\eps} \norm{\RCD{xy}{xy}\eps} r\duxyz \cr
   & \quad + M_7 \int_{K-B_\eps} \norm{\RCD{xy}{xy}\eps} r\duxyz \cr
   & \leq M_8 {(\eps R_0)^3 \over \eps^2} + M_9 \eps^2 \left( {1\over
   \eps R_0} - {1\over R_K} \right) \cr
}$$
and
$$ I_3 = \int_{K'} \left( 2\pi \e^{\gamma(u)} \bigl(
   1-\e^{-\gamma(u)} \bigr) + O(\eps) \right) \Psi(u,0,0,z) \,\du\,\dz
$$ 
Therefore,
$$ \fl \int_{\Real^4} \RCD{xy}{xy}\eps \Psi \duxyz = \int_{K'}
   2\pi \e^{\gamma(u)} \bigl( 1-\e^{-\gamma(u)} \bigr) \Psi(u,0,0,z)
   \,\du\,\dz +O(\eps) $$
and so,
$$ [\RCD{xy}{xy}{}] \approx 2\pi\e^{\gamma(u)}(1-\e^{-\gamma(u)})
   \dirac2(x,y). $$

The distributional contributions, at the axis, of the Riemann tensor
may be used to calculate the distributional energy-momentum tensor
density;
$$ \eqalign{
   [\tT^u{}_u \sqrt{-\tg}] &\approx T^u{}_u\sqrt{-g} - 2\pi
   \e^{\gamma(u)}\bigl(1-\e^{-\gamma(u)}\bigr) \dirac2(x,y) \cr
   [\tT^z{}_z \sqrt{-\tg}] &\approx T^z{}_z\sqrt{-g} - 2\pi
   \e^{\gamma(u)}\bigl(1-\e^{-\gamma(u)}\bigr) \dirac2(x,y) \cr
   [\tT^a{}_b \sqrt{-\tg}] &\approx T^a{}_b\sqrt{-g}
   \qquad\hbox{(Other components)} \cr
} $$
which still represents a pure radiation solution away from the axis,
but whose distributional contributions on the axis are like those of a
cosmic string.

A physical interpretation to these contributions may be made in the
sense of the mass per unit length of the string: Suppose that $V^a$ is
a unit time-like vector in the direction of $\derv{t}$ and that the
distributional and non-distributional matter densities are defined as
$$ \eqalign{ \tilde\varrho &= \tT_{ab} V^a V^b \cr \varrho &= T_{ab}
   V^a V^b \cr} $$
respectively. We may construct a generalised function of $t$ and $z$,
which we shall use to define the mass-per-unit length of the string
via weak equivalence, by
$$ \mu(t,z)= \pbyp{}{z} \int_{C(t,0,z)} \left[ \tilde\varrho
   \sqrt{\tg^\iii} - \varrho \sqrt{g^\iii} \right] \,\dx'\,\dy'\,\dz'
   $$
where $C(t,a,b)$ is the constant $t$ cylinder $a\leq z\leq b$ of an
arbitrary radius and $\tg^\iii$ and $g^\iii$ denote that the
determinants of the induced metrics on the constant $t$ hypersurfaces.

Here $V^a=e^{\gamma} \delta^a_u$ and 
$$ \eqalign{ 
   & \tg^\iii = -\e^{-2\gamma} \left( 1-H_\eps + 2 (2L_\eps-1)
   \e^{2\gamma} \over 1-H_\eps + 2 L_\eps \e^{2\gamma} \right) \tg, \cr
   & g^\iii = -\e^{-2\gamma} g. \cr
}$$
This will imply that $\sqrt{\tg^\iii} \approx \e^{2\gamma}
\sqrt{-\tg}$ and hence,
$$ \eqalign{
   \left[\tilde\varrho \sqrt{\tg^\iii} - \varrho \sqrt{g^\iii}\right]
   &\approx \e^{-3\gamma(u)} (\tg_{au} \tT^a{}_u \sqrt{-\tg}- g_{au} T^a{}_u
   \sqrt{-g} \cr
   &\approx -\e^{-\gamma(u)} \left\{ \bigl( \tT^u{}_u + {x\over r}
   L_\eps \tT^x{}_u + {y\over r} L_\eps \tT^y{}_u \bigr) \sqrt{-\tg}
   \right. \cr
   & \quad \left. - \bigl( T^u{}_u + {x\over r} T^x{}_u + {y\over r}
   T^y{}_u \bigr) \sqrt{-g} \right\} \cr
   &\approx -\e^{-\gamma(u)} \left\{ \tT^u{}_u \sqrt{-\tg} - T^u{}_u
   \sqrt{-g} + P_2 L_\eps + \gamma'/r \right\}  \cr 
}$$
The first term will be weakly equivalent to $2\pi ( 1-\e^{-\gamma(u)})
\dirac2(x,y)$, where as the second term is weakly null, therefore on
integrating we obtain
$$ \mu(u,z) \approx 2\pi \bigl( 1- \e^{-\gamma(u)} \bigr) \eqno(2) $$ 
which implies that the string has a mass per unit length of
$2\pi(1-\e^{-\gamma(u)})$.  This is equal to the angular deficit since
$t=u$ at $r=0$.

\section{Asymptopia and conservation of energy}
A natural question to ask is whether or not there is a conservation of
energy law. At the local level one usually defines this concept in
terms of the contracted Bianchi identities
$$ T^a{}_{b;a} = 0 $$
which certainly hold if the space-time is at least $C^3$-regular. 
If this was interpreted in the context of generalised functions, it
would always hold since the smoothed metric is, by definition,
$C^\infty$.

In the case of radiating space-times, one may formulate the conservation
law in a global sense; that is, whether or not there is mass loss between
the source, in our case the string at $r=0$, and null infinity $\scri$.

In order to enable measurements to be made at null infinity, one must
embed the space-time $(M,g)$ into a compact space-time $(\hat M,\hat g)$
whose boundary $\partial M$ represents the points at infinity. In order
to preserve causal structure under the embedding, one must impose the
condition that $g$ and $\hat g$ are conformally related; that is the
existence of a smooth function $\Omega$ on $\hat M$ such that
$$ {\hat g}_{ab} = \Omega^2 g_{ab} $$
and on $\partial M$, $\Omega=0$ but $\hnabla_a\Omega\neq0$.

Many formalisms already exist for investigating asymptotic behaviour
of four-dimensional space-times (Geroch, 1977). These are very often
suited to axially-symmetric radiation solutions such as Bondi's
radiating metric. A radiating string differs in that the singularity
will itself extend out to infinity, thus making a choice of asymptote
$(\hat M,\hat g)$ problematic. In the case of our pure radiation
solution, which is cylindrically symmetric, we are really measuring
the mass-per-unit length of the string and can expect to measure the
contribution of this piece of string at infinity. This suggests that a
three-dimensional construction of null infinity would be most
appropriate.

Recently a formalism for constructing asymptotes at null infinity for
cylindrically symmetrical radiating space-times was devised by
Ashtekar \etal (1996) in which a symmetry reduction to a
three-dimensional formalism is performed. This formalism was then used
to investigate the asymptotic behaviour of Einstein-Rosen waves.  Our
application of this formalism to our pure radiation solution
$(\psi=0,\gamma=\gamma(t-r))$ will be two-fold; firstly, to reveal to
what extent it is asymptotically flat, and secondly to calculate the
energy flux at null infinity so we can make comparisons with the
matter measured on the string by distributional techniques.

\subsection{Symmetry reduction}
The full four-dimensional metric~(1) may be written in
the form (Ashtekar \etal 1996);
$$ ds^2 = e^{-2\psi} \gg_{ab} dx^a\,dx^b + e^{2\psi} dz^2 $$
where the 3-metric may be expressed in terms of Bondi coordinates
$(u=t-r,r,\phi)$ 
$$ d\sigma^2 = \gg_{ab} dx^a\,dx^b = e^{2\gamma} (-du^2-2du\,\dr) +
   r^2d\phi^2 $$ 
The corresponding energy-momentum tensor for the 3-metric will be
denoted by $\gT_{ab}$ and may be calculated from Einstein tensor, via
analogous three-dimensional field equations
$$ \three{R}_{ab} - \tfrac12 \three{R} \gg_{ab} = \gT_{ab} $$
In general $\gT_{ab}$ will encode the metric coefficient $\psi$ as a
scalar field. (See Ashtekar \etal (1997) for Einstein-Rosen
waves). In our case we have $\psi=0$ so the only non-zero component of
$\gT_{ab}$ is
$$ \gT_{uu}= -{\gamma(u) \over r} $$
In order to examine the structure of null infinity $\scri$, we shall
need to introduce a conformally related unphysical metric
$\hg_{ab}$ 
$$ d{\hat\sigma}^2 = \Omega^2 d\sigma^2 = e^{2\gamma(u)}(-\hr^2 du^2 +
   2 du\, d\hr ) + d\phi^2
   \eqno(3) $$
where $\Omega=\hr$ and $\hr=r^{-1}$.
Here $\scri^+$ is represented by the points $\hr=0$. Since $\gamma(u)$
is arbitrary and smooth, we automatically have that $\hg_{ab}$ is
smooth across $\scri^+$. Moreover the hypersurface $\scri$ will be
null; this is because we may define the normal vector $\hn^a= \hg^{ab}
\hnabla_b\Omega$, and
$$ \hg_{ab} \hn^a \hn^b = \hr^2 e^{-2\gamma(u)} \to 0 \quad \hbox{(as
   $r\to0$)}$$  

\subsection{Asymptotic flatness}
We now can define asymptotic flatness in the three-dimensional sense.
The notation $f\cong g$ will be used to denote that $f$ and $g$ are equal
at the points of $\scri$.
\definition{1 (Ashtekar \etal (1996))}
A 3-dimensional space-time $(M,\gg)$ is said to be asymptotically flat at
null infinity if there is an embedding into a manifold $(\hat M,\hg)$
with a smooth metric and a boundary $\scri$, topologically $S^1\times
\Real$ such that
\item{(i)} $\hat M - \scri$ is diffeomorphic to $M$. \par
\item{(ii)} $\hg_{ab} = \Omega^2 \gg_{ab}$ for some smooth function
  $\Omega$ on $\hat M$ such that $\Omega\cong0$ and
  $\hnabla_a\Omega\not\cong0$. \par
\item{(iii)} $\Omega \gT_{ab}$ is smooth and $\Omega \gT_{ab} \hg^{ab}
  \cong0$. \par
\item{(iv)} If $\hn^a= \hg^{ab} \hnabla_b\Omega$ and $\hV^a$ is any
  smooth vector field on $\hat M$, tangential to $\scri$ then
  $\Omega^{-1} \gT_{ab} \hn^a \hV^b \cong0$ \par
\item{(v)} If $ \hnabla_a\hn^a \cong0$ then $\hn^a$ is complete on 
  $\scri$. \par
\endproclaim

Certainly conditions (i), (ii) and (v) are satisfied by our unphysical
metric~(3). Condition (iii) is also true since $\gT_{ab}$ is
trace-free and
$$ \Omega \gT_{uu} = -\hr^2 \gamma'(u) \cong 0 $$
However condition (iv) is not satisfied for our pure
radiation solution since
$$ \eqalign{ \Omega^{-1} \gT_{ab} \hn^a \hV^b 
   &= \hr^{-1} \gT_{uu}  \hg{}^{u\hr} \hV^u \cr
   &= \gamma'(u)e^{-2\gamma(u)} \hV^u \cr
}$$
The only way to make it vanish would be to force $\gamma$ to be a constant,
thus making the string static.  However the quantity $\Omega^{-1}
\gT_{ab} \hn^a \hV^b$ will still be finite and smooth at $\scri$.
Condition (iv) was introduced in order to guarantee that the energy
flux of matter across $\scri$ remains finite (Ashtekar \etal 1996).
We shall show that, despite this condition not being met, one may
still define a mass at null infinity.

\subsection{Bondi mass aspect and mass at null infinity}
Mass at $\scri$ may be defined from the energy-momentum tensors.
Rather than working with the energy-momentum tensors $\gT_{ab}$ and
$\hT_{ab}$ directly, we shall follow Geroch (1977) and Ashtekar \etal
(1996) and give definitions in terms of the `modified' energy-momentum
tensors
$$ \eqalign{
   \hL_{ab} &= \Omega \three{S}_{ab} = \Omega (\three{R}_{ab} - \tfrac14
   \three{R} \gg_{ab}) \cr
   \hS_{ab} &= \threeh{R}_{ab} - \tfrac14 \threeh{R} \hg_{ab} \cr
}$$ 
The advantage of such tensors is that their algebraic relationship has
the fewest number of terms;
$$ \hL_{ab} = \Omega \hS_{ab} + \hnabla_a \hn_b - \tfrac12 \Omega^{-1}
   \hg_{ab} \hg_{cd} \hn^c \hn^d \eqno(4) $$

We shall denote the pull backs to $\scri$ of $\hg_{ab}$, $\hL_{ab}$
and $\hS_{ab}$ as $\ung{ab}$, $\unL{ab}$ and $\unS{ab}$
respectively. The fact that the pull-back to $\scri$ of
$\hnabla_a\hn_b$ vanishes and that $\scri$ is null will be
sufficient to guarantee that $\ung{ab}$ is a pull-back to $\scri$ of a
positive definite metric on the space $\cal B$ of orbits of $\hn^a$,
which has $S^1$ topology, and hence that $\ung{ab} = \hm_a \hm_b $,
for some 1-form $\hm_a$ on $\scri$.

If it is the case that for any smooth vector field $\hV^a$, $\hL_{ab}
\hn^a \hV^b \cong0$ and $\hS_{ab} \hn^a
\hV^b \cong0$ then $\unL{ab}$ and $\unS{ab}$ have the form
$$ \unL{ab} = \underline{L} \hm_a \hm_b, \quad
   \unS{ab} = \underline{S} \hm_a \hm_b $$

There are sufficient conditions on the energy momentum tensor which
enable $\hL_{ab} \hn^a \hV^b \cong0$ and $\hS_{ab} \hn^a \hV^b
\cong0$ and are weaker than condition (iv) in
definition~1.

\Proposition
Suppose that $T\Omega^{-2}$, ${\hat f}=\Omega^{-2}\hg_{ab} \hn^a\hn^b$
are finite at $\scri$ and that for any $\hV^a$, tangential to $\scri$,
$\gT_{ab} \hn^a \hV^b \cong 0$ then the quantities $\hL_{ab} \hn^a
\hV^b$ and $\hS_{ab} \hn^a \hV^b$ both vanish at $\scri$
\endproclaim

\Proof
The vanishing of $\hL_{ab} \hn^a \hV^b$ is clear from its definition
$$ \eqalignno{
   & \threeh{S}_{ab} = \gT_{ab} - \tfrac12 T \gg_{ab} \cr
   \implies \quad &\Omega^{-1} \hL_{ab} \hn^a \hV^b = \gT_{ab} \hn^a
   \hV^b - \tfrac12 \Omega^{-2} T \hn_a \hV^b \cong 0 & (5) \cr
   \implies \quad &\hL_{ab} \hn^a \hV^b \cong 0 \cr
}
$$
To get $\hS_{ab} \hn^a \hV^b$ to vanish, we should contract
equation~(4) with $\hn^a \hV^b$ to get
$$ \hS_{ab} \hn^a \hV^b = \Omega^{-1} \hL_{ab} \hn^a \hV^b - \tfrac12
   {\hat f} \hn_a \hV^a - \tfrac12 \Omega \hV^a \hnabla_a{\hat f} $$
then apply~(5) and the fact that ${\hat f}$ is finite at $\scri$.
\endproclaim

With the pull-backs in the above forms, we are able to define the
Bondi mass aspect as
$$ {\hat B}\hm_a \cong (\underline{S}-\underline{L}) \hm_a $$
and if we denote a 1-dimensional cross section of $\scri$, to which
$\derv\phi$ is tangential, as $C$ then the mass at $\scri$ for this
cross section may be defined as
$$ E(C) = \oint \bigl(1-\sqrt{2{\hat B}}\bigr) \hm^a \,dS_a $$
So with our solution we obtain a Bondi mass aspect of
$$ {\hat B} \hm_a \cong \tfrac12 e^{-2\gamma} \delta_a^\phi $$
and the mass at $\scri$ for the cross section for $u=\hbox{constant}$
(written as $C_u$) as
$$ \eqalign{ E(C_u)
   &= \int^{2\pi}_0 \left( 1- e^{-\gamma(u)} \right) \,d\phi \cr
   &= 2\pi\bigl(1-e^{-\gamma(u)}\bigr). \cr
}$$
Therefore the energy at infinity agrees with the mass per unit
length at the axis~(2), which implies that there is zero mass loss
between the axis and null infinity.

\begingroup

\bigbreak\vbox{%
\leftline{\bf References}\medbreak
\frenchspacing
\parskip=\smallskipamount
\parindent=0pt

\def\article#1#2#3#4#5#6{#1 #6 #2 {\it#3} {\bf#4} #5 \par}
\def\book#1#2#3#4{#1 #4 {\it#2} (#3) \par}
\def\inbook#1#2#3#4#5#6{#1 #6 {\it#2} in #3 ed.\ #4 (#5)}

\article{Ashtekar~A, Bi\v c\'ak~J and Schmidt~B~G}
{Asymptotic structure of symmetry reduced general relativity} {Phys.\
Rev.\ D} {55}{669--686}{1997}

\article{Clarke~C~J~S, Vickers~J~A and Wilson~J~P}
{Generalised functions and distributional curvature of cosmic strings}
{Class.\ Quantum Grav.}  {13}{2485--2498}{1996}

\article{Colombeau~J~F}
{A multiplication of distributions} {Journal of Mathematical analysis
and applications} {94}{96--115}{1983}

\article{Colombeau~J~F}
{A multiplication of distributions} {Bulletin of the American
Mathematical Society} {23}{251--268}{1990}

\inbook{Geroch~R~P}
       {Asymptotic structure of space-time}
       {Asymptotic structure of space-time}
       {F~P~Es\-pos\-ito and L~Witten}
       {Plenum}{1977}

\article{Geroch~R~P and Traschen~J}
        {Strings and other distributional sources in general
        relativity} {Phys.\ Rev.\ D} {38}{1017--1031}{1987}

\book{Kramer~D, Stephani~H, Herlt~E and MacCallum~M}
     {Exact solutions of Einstein's Field equations} {Cambridge
     University Press} {1980}

\article{Krishna Rao~J}{Cylindrical waves in general
     relativity}{Proc. Nat. Sci. India}{A 30}{439}{1964}

\article{Schwartz~L}
        {Sur l'impossibilit\'e de la multiplication des
        distributions.}  {C. R. Acad. Sci. Paris}{239}{847--848}{1954}

}\endgroup
\bye